
%
%


\magnification=1200
\hsize=28pc
\vsize=43pc
\baselineskip=20pt plus 2pt


\font\fontb=cmbx10 scaled\magstep2


\def\si{\quad }
\def\sii{\qquad}


\def\a{\alpha}
\def\b{\beta}
\def\g{\gamma}
\def\d{\delta}
\def\e{\epsilon}

\def\th{\theta}

\def\k{\kappa}
\def\l{\lambda}
\def\m{\mu}
\def\n{\nu}

\def\s{\sigma}

\def\f{\phi}
\def\vf{\varphi}
\def\c{\chi}

\def\S{\Sigma}

\def\F{\Phi}


\def\pd{\partial}

\def\section{\S}
\def\ds{\displaystyle}


\def\tmu{\vglue .5cm%
         \centerline{\it Department of Physics}
         \centerline{\it Tokyo Metropolitan University}
	 \centerline{\it Minami-Ohsawa, Hachiohji, Tokyo 192-03, Japan}}

\vglue 1.0cm
\centerline{{\fontb
The Improved Bounce Solution }}
\centerline{{\fontb
of SU(2)-Higgs Model}}
\vglue 2.0cm
\centerline{{\bf Yuji Kobayashi }}
\vglue 1.0cm
\tmu
\vglue 1.0cm
\centerline{{\bf ABSTRACT}}
\vglue 0.2cm
We further develop the reduced action formalism
of the SU(2)-Higgs model
originally given by Aoyama et.al..
Our new ansatz for the sphaleron solution makes it possible
to apply this formalism to all range of the Higgs self coupling constant.
Based on the formalism, we construct a bounce solution
oscillating around the sphaleron.
\vfill \eject
The standard model of electroweak interactions
contains the baryon number violation
due to the anomaly [1,2].
Actually the evaluation of the anomalies in this theory
shows that the change of the baryon number
is proportional to the Pontryagin number,
which is the topological number
for the SU(2) gauge field.
Therefore, the baryon number may be changed
by the non-trivial topological configuration
of the SU(2) gauge field. \par
The pure SU(2) gauge theory
has the non-trivial topological solution
which is called instanton [3].
Since the action of the instanton is $ 8\pi^2 / g^2 \approx 190 $
with the SU(2) gauge coupling constant $ g \approx 0.65 $,
the baryon number violating quantum transition
by the instanton
is suppressed by a factor $ \exp(-190) $
and is certainly unimportant. \par
However, at finite temperature,
there may arise the thermally enhanced transition
rather than the quantum tunneling [9].
The SU(2)-Higgs model,
which is the relevant part of the standard model at $ T<T_c $,
has the static unstable finite energy configuration,
which is called the sphaleron [7,8].
The sphaleron has the Chern-Simons number 1/2
and the energy $ 4 \pi v \bar E / g $,
where $v$ is the vacuum expectation value of the Higgs field.
Numerical estimates give
$ \bar E \approx 2.21 \sim 3.85 $
for a wide range of the Higgs self coupling constant
$ \l \approx 0 \sim \infty $.
Since $v$ must vanish at $T_c$,
the thermal transition probability
for the sphaleron
seems to be order unity near $T_c$. \par
The perturbation theories [4,5,6] around the sphaleron
actually shows that
the baryon number violation rate per unit volume
is large [10].
An application to the early universe
leads to the embarrassing result that the thermal transition
for the sphaleron
wipes out the net baryon number
even if it was generated in the grand unification epoch.
So we may need to go into more detail
to resolve the inconsistency problem
between the theory and the observations
that the baryon number density per the photon number density
is of the order of $10^{-10}$.
For example,
the effect of thermal nonequilibrium
has been discussed in ref.[13]. \par
To understand the baryon number violation process
in such cosmological setting,
it may be useful to construct a formalism by which
we can obtain closed expressions of various physical quantities.
The method of reduced quantum mechanical action
proposed by Aoyama, Goldberg, and Ryzak [11] appears
to be a good candidate for such formalism. \par
In this paper, we futher develop the reduced action formalism
so that it is applicable for all range of
the Higgs self coupling constant.
Based on this formalism,
we construct a bounce solution
which is a periodic solution at finite temperature,
and interpolates between the sphaleron and instanton-like configulation. \par
We will see the behaviour of the bounce solution
for all range of Higgs self coupling constant. \par
\vglue 1cm
First, we consider the SU(2)-Higgs model.
It may serve as an effective field theory
to investigate the baryon number violation
in the standard electroweak gauge theory.
SU(2)-Higgs model is given by
$$
S= \int d^4 x \Bigl( {-1 \over 2 g^2} {\rm tr} F_{\m\n} F^{\m\n}
+ \vert D_\m \F \vert ^2 + \l ( \vert \F \vert ^2 -v^2 )^2 \Bigr),
\eqno(1) $$
$$ \eqalign{
&D_\m \F = ( \pd_\m + A_\m ) \F, \cr
&F_{\m\n}=\pd_\m A_\n - \pd_\n A_\m + [ A_\m , A_\n ], \cr}
\eqno(2) $$
where $A_{\m}$ are the SU(2) gauge fields,
$\F$ is the SU(2) doublet Higgs field,
$g \approx 0.65$ is the SU(2) gauge coupling constant,
$\l$ is the Higgs self coupling constant,
and $v$ is the vacuum expectation value of the Higgs field.
$v$ depends on the temperature T as
$$
v=v(T)=v(0) \sqrt{1-{T^2 \over T_c^2}},
\eqno(3) $$
where $v(0) \approx 175 {\rm GeV}$, and
$$
T_c = {2v(0) \over \sqrt{1+{\ds 3 \over \ds 8}{\ds g^2 \over \ds \l}}}
\eqno(4) $$
is the electroweak transition temperature.
$v$ vanishes at and above the electroweak transition temperature
as can be seen in (3). \par
It is well known that the sphaleron solution [7,8]
exists in this theory, and has the following properties;
it is a static, finite-energy but unstable solution
which has the Chern-Simons number $1/2$.
The solution can be written in polar coodinate as
$$
A_0 = 0, \sii A_i = -f \pd_i U U^{-1}, $$
$$
\F = h U \left( \matrix{ 0 \cr v \cr } \right), $$
$$ U = \left( \matrix{ \cos \th & \sin \th \exp (i \vf) \cr
            - \sin \th \exp (-i \vf) & \cos \th \cr } \right),
\eqno(5) $$
where $f$ and $h$ are functions of $r$
which are to be determined by numerical analysis.
Since the energy of the sphaleron is estimated as
$$
E={4 \pi v \over g} \bar E, $$
$$
\bar E \approx 2.21 \sim 3.85 \sii
({\rm for} \si \l =0 \sim \infty ),
\eqno(6) $$
the baryon number violating transition
seems to be unsuppressed near
the electroweak transition temperature. \par
Following Manton [7], we construct the noncontractible loop
of field configurations passing through the sphaleron as
$$ A_0 =0, \sii A_i =-f \pd_i U U^{-1}, $$
$$ U= \left( \matrix{
  \exp (i \m ) (\cos \m -i \sin \m \cos \th )
& \sin \m \sin \th \exp (i \vf ) \cr
  -\sin \m \sin \th \exp (-i \vf )
& \exp (-i \m ) (\cos \m +i\sin \m \cos \th ) \cr
} \right), $$
$$ \F= v \left( \matrix{
h \sin \m \sin \th \exp (i \vf ) \cr
\exp (-i \m ) (\cos \m + ih \sin \m \cos \th ) \cr
} \right) ,
\eqno(7) $$
where $ \m $ is a constant parametrizing the loop.
At $ \m =\pi / 2 $, (7) gives the sphaleron
and it reduces to the vacua at $\m=0$ and $\pi$.
Let us assume that the lowest partial wave of the SU(2)-Higgs model
is relevant for the baryon number violation. Then we can use
an effective two dimensional ($t$ and $r$) formulation,
that is,
$$
A_0 = a_0 {\s_{i} x^{i} \over 2ir}, \sii
A_{i} = {\a e_{i}^{1}+(1+\b)e_{i}^{2}+a_{1}re_{i}^{3} \over 2ir},
$$
$$
\F = (\g + i\d {\s_{i} x^{i} \over r})
     \left( \matrix{ 0 \cr v \cr } \right),
\eqno(8)$$
where
$$
e_{i}^{1}=\s_{i}-{\s_{j}x^{j}x_{i} \over r^2}, \sii
e_{i}^{2}=i{\s_{j}x^{j}\s_{i}-x_{i} \over r}, \sii
e_{i}^{3}={\s_{j}x^{j}x_i \over r^2}.
\eqno(9)$$
$a_\m$ are two dimensional vector fields,
$\c=\a+i\b$ and $\f=\g+i\d$ are the complex scalar fields
on the two dimensional space-time.
The action of the SU(2)-Higgs model then reduces to
$$ \eqalign{
S &= \int dtdr \si 4 \pi r^2 \cdot \cr
  \biggl( &\si {1 \over g^2 } \Bigl({1 \over 2}(\e^{\m\n}\pd_\m a_\n )^2
    +{1 \over r^2 } \vert D_\m \c \vert ^2
    +{1 \over 2r^4 }(\vert \c \vert ^2 -1)^2 \Bigr) \cr
  &+v^2 \Bigl( \vert D_\m \f \vert ^2 +{1 \over 2r^2 } \vert \f \vert ^2
    +{1 \over 2r^2 }\vert \c \vert ^2 \vert \f \vert ^2
    -{1 \over r^2 }{\rm Im}(\c^* \f^2 )\Bigr) \cr
  &+\l v^4 (\vert \f \vert ^2 -1)^2 \biggr), \cr }
\eqno(10) $$
$$ \eqalign{
D_\m \c =(\pd_\m -ia_\m ) \c, \cr
D_\m \f =(\pd_\m -{i \over 2} a_\m ) \f. \cr }
\eqno(11) $$
The noncontractible loop (7) can be rewritten
in the two dimensional formulation as
$$ \eqalign{
a_0 &=0, \sii a_1 =-(2\m -\pi ) \pd_r h, \cr
\c &=-i \exp \Bigl( -i(2\m -\pi ) h \Bigr)
\biggl( f \Bigl( \exp (2i\m )-1 \Bigr) +1 \biggr), \cr
\f &=\exp \Bigl(-{i \over 2}(2\m -\pi ) h \Bigr)
(\cos \m +ih\sin \m ), \cr }
\eqno(12) $$
where we have used the global SU(2) symmetry
and the custodial symmetry
of the SU(2)-Higgs model
in reducing the system into the two-dimensional formulation.
We have also used the U(1) symmetry
in the two-dimensional action (10)
to eliminate the $\m$ dependence in the asymptotic region
($r \to \infty $). \par
Now we regard the system as quantum mechanical one
by elevating the parameter
$ \m $ to the time-dependent dynamical variable.
For the sake of analytical estimates,
we employ the functions $f$ and $h$ of the following forms;
$$ \eqalign{
f&=\Bigl( 1-\exp (-{gvr \over a }) \Bigr)^2, \cr
h&=1-\exp \Bigl(-(2\sqrt{2l}+1){gvr \over a }\Bigr), \cr }
\eqno(13) $$
where $l=\l /g^2$.
We have introduced the variational parameter $a$
to be determined by minimizing the action.
This form is different from that given by
Aoyama et.al.[11].
The essential difference between our ansatz (13)
and the Aoyama et.al.'s exists in that ours uses
the sphaleron as the boundary condition
whereas the Aoyama et.al.'s is motivated by
and is close to the instanton solution. \par
Our ansatz also differs from that given by Funakubo et.al.[14].
The difference exists in that ours has $\l$ dependence which the sphaleron must
have,
while Funakubo et.al.'s does not have. \par
As a result of these feature, our ansatz can be applicable for any $\l$. \par
Substituting (12) with (13) to the action (10), we obtain
the reduced action as
$$ S_{{\rm red}}={4\pi v \over g}
\int^{1 \over T}_0 dt \Bigl({M \over 2}({d \m \over dt })^2 +V \Bigr)
\eqno(14) $$
at finite temperature $T$, where
$$ \eqalign{
M&={1 \over (gv)^2}(A+B \sin ^2 \m), \cr
V&=C \sin ^2 \m + D \sin ^4 \m . \cr}
\eqno(15) $$
The parameters $ A,B,C,$ and $D$ are given as follows:
$$ \eqalign{
A&={22 a \over 3} + {16 a \over 3+2 \sqrt{2l}}
  + {16 a \over -1- \sqrt{2l}}
  + {5 a \over 1+2 \sqrt{2l}}, \cr
B&= {128 a \over -5-4 \sqrt{2l}}
  + {64 a \over -3-4 \sqrt{2l}}
  + {32 a \over -5-2 \sqrt{2l}} \cr
 &+ {144 a \over -3-2 \sqrt{2l}}
  + {64 a \over 2+ \sqrt{2l}}
  + {72 a \over 1+ \sqrt{2l}}
  + {635 a^3 \over 432(1+2 \sqrt{2l})^3}, \cr
C&= {4 \over 3 a} + {11 a \over 6}
  + {4 a \over 3+2 \sqrt{2l}}
  + {4 a \over -1- \sqrt{2l}}
  + {5 a \over 4 (1+ \sqrt{2l})}, \cr
D&= {8 k \over a}
  + {8 a \over -5-4 \sqrt{2l}}
  + {4 a \over -5-2 \sqrt{2l}}
  + {19 a \over -3-2 \sqrt{2l}} \cr
 &+ {8 a \over 2+ \sqrt{2l}}
  + {6 a \over 1+ \sqrt{2l}}
  + { a \over -1-2 \sqrt{2l}}
  + {635 a^3 l \over 864 (1+ \sqrt{2l})^3}, \cr }
\eqno(16) $$
where
$$
k=516 \ln 2 + 96 \ln 3 - 220 \ln 5 - 56 \ln 7 \approx 0.0834.
\eqno(17) $$
The quantity $k$ arises from the subtle space integration
of the self coupling term of $\c$ in (10).
The classical solution
corresponding to the sphaleron is $\m=\pi /2$
with the action
$$ S_{{\rm sp}}={4\pi v \over g}{ \bar E_{{\rm sp}} \over T},
\eqno(18) $$
where
$$ \bar E_{{\rm sp}}=C+D={X \over a} + Y a + Z a^3
\eqno(19) $$
with
$$ \eqalign{
X &= {4 \over 3} + 8k, \cr
Y &= {11 \over 6} + {8 \over -5-4 \sqrt{2l}}
      + {4 \over -5-2 \sqrt{2l}} + {15 \over -3-2 \sqrt{2l}} \cr
   &+ {8 \over 2+ \sqrt{2l}} + {1 \over -1-2 \sqrt{2l}}
      + {13 \over 4(1+ \sqrt{2l})}, \cr
Z &= {635 l \over 864 (1+2 \sqrt{2l})^3}. \cr }
\eqno(20) $$
{}From (19), $ \bar E_{{\rm sp}} $ takes the minimum with
$$ \eqalign{
a &= \sqrt{{-Y + \sqrt{Y^2 + 12 X Z } \over 6 Z }} \cr
  &= 4 \sqrt{{5(1+6k) \over 41}} \sim
     2 \sqrt{{2(1+6k) \over 11}} \cr
  &\approx 1.71 \sim 1.04 \sii ({\rm for} \si l=0 \sim \infty ), \cr }
\eqno(21) $$
and the minimum value of $\bar E_{{\rm sp}}$ is given as
$$ \eqalign{
   \bar E^{{\rm min}}_{{\rm sp}}
    &={2 \over 3} \sqrt{{41 (1+6k) \over 5}} \sim
      {2 \over 3} \sqrt{22 (1+6k)} \cr
    &\approx 2.34 \sim 3.83 \sii
({\rm for} \si l=0 \sim \infty ). \cr }
\eqno(22) $$
This reproduces well the energy of the sphaleron
even when $ \l \to \infty $.
This feature is in sharp contrast with those of
Aoyama et.al.'s and Funakubo et.al.'s cases. \par
Next, we investigate the bounce solution.
For the sake of analytical estimate,
we approximate $M$ and $V$ in the action as
$$ \eqalign{
 M &\to {1 \over (gv)^2}(A+B), \cr
 V &\to (C+D) \sin^2 \m. \cr
}\eqno(23) $$
In fact, the approximation is a rather good one;
if we do not use this approximation
strictly keeping (15),
the result is different only by $ 6 \sim 17 \% $
as will be shown later.
Under this approximation, the action reduces to that of a pendulum
and the classical periodic solution exists:
$$
\m =\arccos \Bigl( -\k {\rm sn}
     (\sqrt{2 (C+D) \over A+B } gvt;\k) \Bigr),
\eqno(24) $$
where $\k$ is the modulus which takes value from $0$ to $1$.
This solution may be called as bounce solution
because this solution represents the motion
swinging around the sphaleron
in the one dimensional space.
In order that this solution has the period $1/T$,
the following equation must be satisfied:
$$
\sqrt{2 (C+D) \over A+B }{gv \over T}=4K(\k),
\eqno(25) $$
where $K(\k)$ is the elliptic integral of the first kind
given by
$$
K(\k)=\int^1_0 dx {1 \over \sqrt{(1-x^2)(1-\k ^2 x^2)}},
\eqno(26) $$
which varies from $\pi /2$ to $\infty$ as
$\k$ increases from $0$ to $1$.
Notice that the equation (22) relates
the modulus $\k$ and the temperature $T$.
The action of the bounce solution becomes
$$
S_\k ={16 \pi \over g^2} \sqrt{(A+B)(C+D) \over 2}
\Bigl( 2 E(\k)-(1-\k^2)K(\k) \Bigr),
\eqno(27) $$
where $E(\k)$ is the elliptic integral of the second kind
given as
$$
E(\k)=\int^1_0 dx \sqrt{ 1-\k^2 x^2 \over 1-x^2 },
\eqno(28) $$
which varies from $\pi/2$ to $1$ as
$\k$ increases from $0$ to $1$.
Now we consider two extreme cases of the bounce solution.
When $ \k =0 $, the temperature and the action take the form
$$ \eqalign{
T &= {gv \over 2 \pi } \sqrt{2(C+D) \over A+B }, \cr
S_0 &= { 8\pi^2  \over g^2} \sqrt{(A+B) (C+D) \over 2}
 ={4 \pi v \over g} {\bar E_{{\rm sp}} \over T}, \cr }
\eqno(29) $$
The bounce solution reduces to the sphaleron at $\k=0$.
Similarly, when $ \k =1 $,
they become
$$ \eqalign{
T &=0, \cr
S_1 &= {16\pi^2 \over g^2} \s, \cr }
\eqno(30) $$
where
$$
\s={1 \over \pi}\sqrt{2(A+B)(C+D)}.
\eqno(31) $$
Since $\s$ takes minimum at $a=0$,
the minimum value is given as
$$ \eqalign{
\s_{{\rm min}} &={2 \over 3\pi}\sqrt{26(1+6k)} \sim
     {4 \over 3\pi}\sqrt{11(1+6k)} \cr
   &=1.33 \sim 1.73 \sii ({\rm for} \si l=0 \sim \infty). \cr }
\eqno(32) $$
We call the bounce solution at $\k =1$
as instanton-like solution.
Its action has an expression similar to the instanton
in the pure SU(2) gauge theory.
If we work with (15)
without the approximation (23),
we obtain $\s_{{\rm min}} \approx 1.11 \sim 1.62$
for $l=0 \sim \infty $.
Then the solution is even closer to the instanton. \par
Therefore, we recognize that
the bounce solution interpolates between
the sphaleron and the instanton-like solution
as shown in Fig.1 $\sim$ 3.
Fig.1,2 and 3 show the cases of $l$=0.1,1 and $\infty$, respectively.
It should be noticed that
the sphaleron dominant phase may directry change to
the instanton dominant phase
since the action of the bounce solution is larger than
that of the sphaleron, or the instanton-like
at any temperature and at any $\l$.
This is in contrast to the case of O(3) non-liner sigma model [12].
The instanton-like solution transits to the sphaleron
at the temperature
where the action of the instanton-like
and that of the sphaleron is equal.
The temperature correponding to this transition
is given by
$$
T_0 ={2v(0) \over
\sqrt{
1+{\ds 3 \over \ds 8}{\ds g^2 \over \ds \l}
 +{\ds 64 \pi^2 \s^2_{{\rm min}} \over
       \ds g^2 (\bar E^{{\rm min}}_{{\rm sp}})^2}
}
}.
\eqno(33) $$
Naturally this is lower
than the electroweak transition temperature $T_c$
in the all range of $\l$.
The baryon number violating transition by the sphaleron
occurs in the temperature range
$ T_0 < T < T_c $. \par
\vglue 1cm
We have provided a bounce solution interpolating between
instanton-dominated and sphaleron-dominated processes
for all range of $\l$. \par
It seems to be important applying this new bounce solution
to the former calculations which employ the bounce solutions
(for example, [14])
since this new one is applicable for all range of $\l$. \par
\vglue .5cm
\centerline{\bf Acknowledgements}
\vglue .5cm

The auther would like to thank Professor H.Minakata for useful comments.

\vfill \eject
\centerline{{\bf References}}
\vglue .5cm
\item{[1]} J.S.Bell and R.Jackiw,
Nuovo Cimento 51, 47 (1969)

\item{[2]} S.L.Adler,
Phys. Rev. 177, 2426 (1969)

\item{[3]} G.'t Hooft,
Phys. Rev. D14, 3432 (1976)

\item{[4]} I.Affleck,
Phys. Rev. Lett. 46, 388 (1981)

\item{[5]} A.Linde,
Nucl. Phys. B216, 421 (1981)

\item{[6]} Mottla,
Nucl. Phys. B203, 581 (1982)

\item{[7]} N.Manton,
Phys. Rev. D28, 2019 (1983)

\item{[8]} F.Klinkhamer and N.Manton,
ibid. 30, 2212 (1984)

\item{[9]} V.Kuzmin, V.Rubakov and M.Shaposhnikov,
Phys.Lett.B155,36 (1985)

\item{[10]} P.Arnold and L.McLerran,
Phys. Rev. Lett. D36, 581 (1987)

\item{[11]} H.Aoyama, H.Goldberg and Z.Ryzak,
Phys. Rev. Lett. 60, 1902 (1988)

\item{[12]} K.Funakubo, S.Otsuki and F.Toyoda,
Prog.Theor.Phys.84, (1990)

\item{[13]} L.McLerran, M.Shaposhnikov, N.Turok and M.Voloshin,
Phys. Lett. B256, 451 (1991)

\item{[14]} K.Funakubo, S.Otsuki, K.Takenaga and F.Toyoda,
Prog.Theor.Phys.89, 881 (1993)

\bye